\documentclass[10pt,journal,compsoc]{IEEEtran}
\makeatletter
\long\def\@makecaption#1#2{\ifx\@captype\@IEEEtablestring%
\footnotesize\begin{center}{\normalfont\footnotesize #1}\\
{\normalfont\footnotesize\scshape #2}\end{center}%
\@IEEEtablecaptionsepspace
\else
\@IEEEfigurecaptionsepspace
\setbox\@tempboxa\hbox{\normalfont\footnotesize {#1.}~~ #2}%
\ifdim \wd\@tempboxa >\hsize%
\setbox\@tempboxa\hbox{\normalfont\footnotesize {#1.}~~ }%
\parbox[t]{\hsize}{\normalfont\footnotesize \noindent\unhbox\@tempboxa#2}%
\else
\hbox to\hsize{\normalfont\footnotesize\hfil\box\@tempboxa\hfil}\fi\fi}
\makeatother

\usepackage[utf8]{inputenc}
\ifCLASSOPTIONcompsoc
  \usepackage[nocompress]{cite}
\else
  \usepackage{cite}
\fi

\usepackage{amsfonts}

\usepackage[pdftex]{graphicx}
\graphicspath{{./images/}}
\DeclareGraphicsExtensions{.pdf,.jpeg,.png,.svg}
\usepackage{amsmath}
\usepackage{xspace}
\usepackage{multirow}
\usepackage{amssymb}
\usepackage{algorithm}
\usepackage[noend]{algpseudocode}
\renewcommand{\algorithmicrequire}{\textbf{Input:~}}
\renewcommand{\algorithmicensure}{\textbf{Output:~}}

\usepackage{array}
\usepackage{color}
\usepackage{listings}
\usepackage{placeins}
\usepackage{enumitem}

\algnewcommand{\algorithmicforeach}{\textbf{for each}}
\algdef{SE}[FOR]{ForEach}{EndForEach}[1]
  {\algorithmicforeach\ #1\ \algorithmicdo}
  {\algorithmicend\ \algorithmicforeach}

\newcommand\acro{{\sc{INFuseR-MG\xspace}\xspace}\xspace}

\usepackage[caption=false,font=normalsize,labelfont=sf,textfont=sf]{subfig}

\begin{document}
\title{Boosting Parallel Influence-Maximization Kernels for Undirected Networks with Fusing and Vectorization}

\author{G\"{o}khan~G\"{o}kt\"{u}rk
        and~Kamer~Kaya
\IEEEcompsocitemizethanks{\IEEEcompsocthanksitem G. G\"{o}kt\"{u}rk and K. Kaya are with Computer Science and Engineering, Faculty of Engineering and Natural Sciences, Sabanci University, Istanbul, Turkey.}
}
\markboth{}%
{G\"{o}kt\"{u}rk \MakeLowercase{\textit{et al.}}: Boosting Parallel Influence-Maximization Kernels for Undirected Networks with Fusing and Vectorization}

\maketitle

\begin{abstract}
Influence maximization~(IM) is the problem of finding a seed vertex set which is expected to incur the maximum influence spread on a graph. It has various applications in practice such as devising an effective and efficient approach to disseminate information, news or ad within a social network. The problem is shown to be NP-hard and approximation algorithms with provable quality guarantees exist in the literature. However, these algorithms are computationally expensive even for medium-scaled graphs. Furthermore, graph algorithms usually suffer from spatial and temporal irregularities during memory accesses, and this adds an extra cost on top of the already expensive IM kernels. In this work, we leverage fused sampling, memoization, and vectorization to restructure, parallelize and boost their performance on undirected networks. The proposed approach employs a pseudo-random function and performs multiple Monte-Carlo simulations in parallel to exploit the SIMD lanes effectively and efficiently. Besides, it significantly reduces the number of edge traversals, hence the amount of data brought from the memory, which is critical for almost all memory-bound graph kernels. We apply the proposed approach to the traditional {\sc MixGreedy} algorithm and propose \acro{} which is more than $3000\times$ faster than the traditional greedy approaches and can run on large graphs that have been considered as {\em too large} in the literature. For instance, the new algorithm runs in $2.09$, $0.08$, $0.36$ seconds on graphs {\tt Amazon}, {\tt NetHEP}, {\tt NetPhy} with 16 threads where the sequential baseline takes $141.3$, $259.1$ and $1725.2$ seconds, respectively. To compare \acro{} with the state-of-the-art approximation algorithms, we conduct a thorough experimental analysis with various influence settings. The results on real-life, undirected networks show that on 16 threads, \acro{} is $2.3\times$--$173.8\times$ faster than state-of-the-art while being superior in terms of influence scores, and using a comparable amount of memory.
\end{abstract}

\IEEEpeerreviewmaketitle

\section{Introduction}
With their rapid growth, the study of effective information diffusion in networks becomes a fruitful area of research with several applications from many fields such as viral marketing~\cite{leskovec2007dynamics,trusov2009effects}, social media analysis~\cite{zeng2010social, moreno2004dynamics}, and recommendation systems~\cite{lu2012recommender}. Since these networks have been used for educational, political, economical, and social purposes, the diffused information can have various importance levels. Furthermore, the diffusion can be a time-critical process, but it can be costly to increase its speed and coverage by other means. Hence, novel approaches to find good vertex sets which effectively spreads information are vital in practice.\looseness=-1

The Influence Maximization~(IM) problem is introduced by Kempe~et~al.~\cite{kempe2003maximizing}. Formally, it focuses on finding the most promising seed~(vertex) set with a given cardinality that increases the expected number of influenced vertices. 
IM is proven to be NP-hard \cite{kempe2003maximizing} and there are various simplifications and heuristics proposed in the literature~\cite{MixGreedy, narayanam2010shapley, kimura2007extracting, chen2010PMIA,chen2010LDAG, kim2013scalable, goyal2011simpath, jung2012irie,cheng2014imrank,liu2014influence,galhotra2016holistic}. It has also been shown that a greedy Monte-Carlo approach provides a constant approximation for the optimal solution~\cite{kempe2003maximizing}. 
For a graph with $n$ vertices, the expected complexity of this greedy algorithm, estimating an influence score $\sigma$, running $R$ simulations and selecting $K$ seed vertices is $\mathcal{O}(KRn\sigma)$.
Hence, for real-life networks with hundreds of thousands of vertices, the approach is expensive. However, these simulation-based, greedy algorithms provide the best possible approximation guarantees. Therefore they are considered as the gold standard for IM.  

Performing the simulations of a greedy algorithm in parallel is an immediate and straightforward remedy to reduce the execution time of IM kernels and make them scalable for large-scale networks. However, restructuring the kernels to leverage instruction-level parallelism have not been investigated before. Although modern compilers can efficiently and automatically utilize instruction-level parallelism for applications with regular memory access patterns, it is not a straightforward task for graph processing kernels due to their irregular memory accesses. Furthermore, vectorization attempts on such kernels usually fail to provide significant performance improvements. In this work;
\begin{itemize}[leftmargin=*]
\item We propose \acro{}, an ultra-fast and high-quality Influence Maximization algorithm for undirected networks. Unlike the traditional greedy approach, the proposed approach samples the edges as they are being traversed in multiple simulations. Hence, for a single simulation, sampling and diffusion processes are {\em fused}.

\item By running concurrent simulations at once, we reduce the amount of connectivity information read from the memory. 
Hence, the proposed approach {\em reduces the pressure on the memory sub-system}.
Furthermore, we utilize vectorized instructions almost with full efficiency for the {\em cascade} model to {\em regularize the memory-accesses}.

\item \acro can be around $200000\times$ faster compared to the traditional greedy approaches. It is usable on large graphs that have been considered as {\em too large} in the literature. For instance, the new algorithm runs in $2.09$, $0.08$, $0.36$ seconds on networks {\tt Amazon}, {\tt NetHEP}, {\tt NetPhy} with 16 threads where the sequential baseline takes $141.3$, $259.1$ and $1725.2$ seconds, respectively. In fact, with a 302,400 seconds~(3.5 days) timeout, the sequential baseline can process only the above-mentioned 3 (out of 12) real-life graphs, having 1.2M, 58.9K and 231.5K edges. On the other graphs, the original algorithm cannot complete the simulations within the time limit. However, \acro{} completes all of 12 graphs around 1200 seconds in total, where the maximum is 654 seconds for the {\tt Orkut} network having 3.1M vertices and 117.2M edges. 

\item To better position the performance of \acro{} in the IM literature, we compare the performance, memory usage and influence score with a state-of-the-art approximation algorithm {\sc Imm}~\cite{minutoli2019fast}. The experiments show that \acro{} is $2.3\times$--$173.8\times$ faster than state-of-the-art while always being (marginally) superior in terms of influence scores, and using a comparable amount of memory. To be fair, we want to emphasize that the state-of-the-art tool can also work with directed graphs where \acro{} only supports undirected graphs. 
\end{itemize}

The paper is organized as follows: 
In Section~\ref{sec:notation}, we present 
the background on IM and introduce the mathematical notation. 
Section~\ref{sec:infuser} describes the proposed approach in detail.
In Section~\ref{sec:exp}, a thorough performance comparison over the traditional algorithms is provided by conducting experiments on various real-world datasets and influence settings. Besides, a comparison with the state-of-the-art from the literature is given.
Section~\ref{sec:rel} presents a comparative overview of the existing work. Finally, Section~\ref{sec:conc} discusses future work and concludes the paper.

\section{Notation and Background}\label{sec:notation}
Let $G = (V,E)$ be an undirected graph where the $n$ vertices in $V$ correspond the agents, and $m$ edges in $E$ correspond the relations between the agents in $V$.
The neighborhood of a vertex $u \in V$ is denoted as $\Gamma_G(u) = \{v: \{u,v\} \in E\}$.
Each edge $\{u, v\} \in G$
has a weight $w_{u,v}$ associated with the diffusion probability from $u$ to $v$. In practice, $w_{u,v}$ can be determined by the strength of $u$ and $v$'s relationship. 
Although the graph is undirected, to emphasize the direction of diffusion, we will use tuples instead of sets to denote edges. That is an edge $\{u,v\} \in E$ can be encountered either in form $(u, v)$~($u$ influences $v$) or $(v, u)$~($v$ influences $u$). 
 
A graph $G' = (V',E')$ is a subgraph of $G$ if $V' \subseteq V$ and $E' \subseteq E$. If all the vertices in $G'$ are connected $G'$ is called as a {\em connected subgraph} of $G$. If the subgraph is a maximally connected subgraph of $G$ it is called a {\em connected component}~(CC) of $G$. 

\begin{table}[!ht]
    \caption{Table of notations}
    \label{tab:notation}
    \centering
    \begin{tabular}{|l|p{0.74\linewidth}|}
        \hline
        Variable & Definition  \\
        \hline
        $G = (V,E)$     & Graph $G$ with vertices $V$ and edges $E$ \\
        $\Gamma_G(v)$   & Neighborhood of vertex $v$ in graph $G$\\
        $w_{u,v}$       & Probability of $u$ directly influencing $v$ \\
        $R_{G}(v)$     & Reachability set of vertex $v$ on graph $G$\\
        \hline\hline
        $S$             & Seed set to maximize influence\\
        $K$             & Size of the seed set\\
        $\mathcal{R}$   & Number of Monte-Carlo simulations performed\\
        $\sigma_{G}(S)$& Influence score of $S$ in $G$, i.e., expected number of vertices reached from $S$ in $G$\\
        $\sigma_{G}{(S,v)}$          & Marginal influence gain by adding vertex $v$ to seed set $S$\\
\hline\hline
        $h(u,v)$        & Hash function for edge $\{u,v\}$\\
        $h_{max}$       & Maximum value hash function $h$ can return\\
        \hline\hline
        $B$ & Batch size, number of simultaneous simulations ran.\\
        $[a, \ldots, a]_B$      & Vector of size $B$, contains all $a$\\
        \hline         
    \end{tabular}
\end{table}

\subsection{Influence Maximization}

\begin{figure}[!ht] 
    \centering
  \subfloat[\small{IC}\label{fig:ic}]{%
       \includegraphics[width=0.45\linewidth]{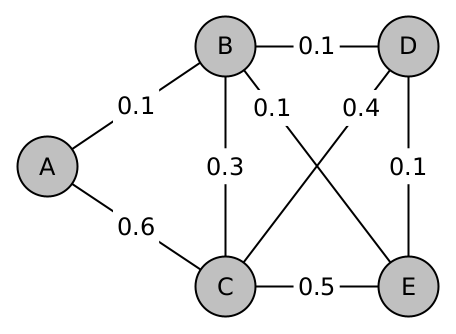}}
  \subfloat[\small{WC}\label{fig:wc}]{%
        \includegraphics[width=0.45\linewidth]{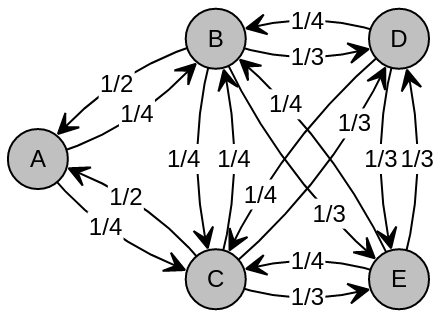}}
    \\
  \caption{\protect\subref{fig:ic} 
The undirected graph $G = (V, E)$ for Independent Cascade with independent diffusion probabilities.
\protect\subref{fig:wc}
The directed graph obtained from the undirected one by setting the diffusion probabilities of incoming edges to $1 / |\Gamma_G(v)|$ for each vertex $v \in V$. 
  }
  \label{fig:xx} 
\end{figure}

Given a graph, Influence Maximization aims to find a seed set $S \subseteq V$ among all possible size $K$ subsets of $V$ that maximizes an {\em influence spread function} $\sigma$  when the diffusion process is initiated from $S$. Although we focus on undirected graphs, for IM, the graph can be directed or undirected depending on the initial construction. Figure~\ref{fig:xx} shows a~(Fig.~\ref{fig:ic}) and directed~(Fig.~\ref{fig:wc})
graph for which the weights on the edges are diffusion/influence probabilities. 

The influence spread function $\sigma_{G,M}(\cdot)$ computes the {\em expected} number of agents/nodes/vertices influenced~(activated) through a diffusion model $M$. For the sake of simplicity, we drop $M$ from the notation; in the rest of the text, $\sigma_{G}$ refers to $\sigma_{G,M}$. Some of the popular diffusion models for IM in the literature are {\em independent} and {\em weighted cascade}~(IC and WC),  and {\em linear threshold}~(LT)~\cite{kempe2003maximizing}. 

\begin{itemize}[leftmargin=*]
\item {\bf Cascade} model runs in rounds and activates a vertex $v$ if one of its (incoming) edges, $(u,v)$, is used during the diffusion process which happens with probability $w_{u, v}$ given that $u$ have already been activated in the previous rounds. In the {\em independent} variant, which we focus on in this work, activation probabilities are independent (from each other and previous activations) as in Figure~\ref{fig:ic}. 
The {\em weighted} variant of the cascade model uses a directed graph $G$ as in Figure~\ref{fig:wc} (even when the original graph is undirected). 
A classical approach for assigning the edge weights, as performed in~\cite{MixGreedy}, 
is setting $w_{u,v} = 1 / d_v$ where $d_v$ is the number of incoming edges of $v$ (which is equal to $\Gamma_G(v)$ in the original graph). Hence, if $v$ has $\ell$ neighbours activated after the last round, its activation probability in the current round is $1 - (1 - 1/d_v)^\ell$. 


\item{\bf Linear threshold} generalizes the cascade models and activates a vertex $v$ if the total activation coming from $v$'s neighbors surpasses a threshold $\theta_v$. Throughout the process, all the $\{u, v\}$~(or $(u,v)$ in a directed graph) edges with active $u$ vertices are taken into account. When the sum of these edge weights exceeds $\theta_v$, $v$ is activated~\cite{kempe2003maximizing}. 


\end{itemize}


In this paper, we focus on the independent cascade model, but the proposed techniques are also applicable to the other models in the literature for undirected graphs.

\subsection{Existing approaches for Influence Maximization}\label{sec:existing}

There exist simulation-based~\cite{kempe2003maximizing, MixGreedy}, sketch-based ~\cite{SKIM}, and proxy-based~\cite{MixGreedy, jung2012irie} approaches in the literature to find a seed set $S$ that maximizes the influence spread in a graph. Simulation-based approaches run Monte-Carlo simulations whereas sketch-based ones utilize approximate data structures. On the other hand, proxy-based approaches simplify the IM problem and utilize simpler heuristics.

As stated before, the IM problem is NP-hard under the cascade and linear threshold models. The influence function is monotone and submodular, which means that adding a single vertex to the current seed set can only increase the overall influence and decreases the marginal influence scores for the remaining vertices that are not in the set. Due to these properties, the influence score of a greedy solution which always adds the most promising vertex with the highest marginal gain to a seed set of final size $K$ is at least $1-(1-1/K)^K \geq 63\%$ of the optimal solution~\cite{nemhauser1978analysis}. 

Kempe et al.\cite{kempe2003maximizing} proposed the greedy Monte-Carlo-based algorithm using the above-mentioned approach and set the foundations. At each step, the greedy algorithm finds the vertex that increases the influence the most. As Feige's optimal inapproximability result shows~\cite{feige1998threshold}, the guaranteed approximation ratio is the {\em gold standard} for problems with a non-trivial size, both asymptotically and practically. On the contrary, the other, sketch- and proxy-based approaches do not guarantee this approximation ratio. This is why we target the greedy, simulation-based algorithms that use the proposed techniques to boost their performance. To the best of our knowledge, these algorithms have experimented only on small-scale graphs in the literature.

 Since IM is an expensive problem, there exist studies in the literature focusing on improving the algorithmic complexity. Instead of trying all the vertices at each step, the Cost-effective Lazy Forward~(CELF) algorithm of Leskovec et al.~\cite{CELF} keeps the vertices in a priority queue w.r.t. their marginal influence gains. Due to the submodularity property of the influence spread function, these values set upper bounds for the current marginal gains. When a vertex is visited, its current marginal gain is updated, i.e., its exact value is computed, and the vertex is replaced further down in the queue. When a vertex is seen twice, the remaining vertices are guaranteed to have smaller marginal gains. Hence, the greedy decision can be immediately taken. The bottleneck of CELF is in its initialization; the (marginal) influence scores for all the vertices must be computed, which is the most time-consuming part and makes the approach expensive for large-scale graphs. This approach is improved by Goyal~et~al.~\cite{Goyal11} by further exploiting the submodularity of the influence spread function.

Chen et al. improve CELF with {\sc MixGreedy}~\cite{MixGreedy}. Instead of running Monte-Carlo simulations from each vertex to find the initial marginal gains, {\sc MixGreedy} uses one iteration of another IM algorithm {\sc NewGreedy} whose pseudocode is given in Algorithm~\ref{algo:newgredy}. The algorithm greedily chooses $K$ vertices to form the seed set $S$. To choose each seed vertex, $\cal R$ graph samples are used. For each inner-iteration~(lines~\ref{ln:inner_start}--\ref{ln:inner_end}), the algorithm samples a subgraph $G'$ from $G$. The pseudocode of the sampling algorithm, {\sc Sample}, is given in Algorithm~\ref{alg:sample} where each edge $\{u, v\}$ is included with probability $w_{u,v}$. Then the marginal gains of $G'$'s vertices are computed by using the reachability sets. 

\begin{algorithm}
\caption{\sc{NewGreedy}($G,K,\mathcal{R}$)}
\label{algo:newgredy}
\algorithmicrequire{$G = (V,E)$: the influence graph
\\\hspace*{6.6ex}{$K$: number of seed vertices
\\\hspace*{6.7ex}$\mathcal{R}$: number of MC simulations per seed vertex}\\}
\algorithmicensure{$S$: a seed set that maximizes influence on $G$
\\\hspace*{8.6ex}{$mg$: marginal influence scores}}
\begin{algorithmic}[1]
    \State {$S \leftarrow \emptyset$}
    \For{$k=1\ldots K$}
        \For{$v\in V$}
            \State $mg_v\leftarrow 0$
        \EndFor
        \For{$r=1\ldots \mathcal{R}$}\label{ln:inner}
            \State $G' = (V, E') \leftarrow$ {\sc Sample}($G$)\label{ln:inner_start}
            \State Compute $R_{G'}(S)$
            \State Compute $|R_{G'}({\{v\})}|$ for all $v \in V$
            \For{ $v\in V \setminus S$}
                \If{$v \notin R_{G'}(S)$}
                    \State $\sigma_{G'}(S, v) \leftarrow |R_{G'}({v})|$
                    \State $mg_v \leftarrow mg_v + \sigma_{G'}(S, v) $\label{ln:inner_end}
                \EndIf
            \EndFor
        \EndFor
        \State $mg_v \leftarrow \frac{mg_v}{R}$ for all $v\in V \setminus S$
        \State $S \leftarrow S \cup \{{\tt argmax}_{v\in V}\{mg_v\} \} $
        \EndFor
    \State \Return $S$, $mg$
\end{algorithmic}
\end{algorithm}

\begin{algorithm}[!ht]
\caption{\sc{Sample}($G$)}
\label{alg:sample}
\algorithmicrequire{$G = (V, E)$: the original graph}\\
\algorithmicensure{$G' = (V, E')$: a subgraph of $G$}

\begin{algorithmic}[1]
    \State $E' \leftarrow \emptyset$
    \For{each $\{u, v\}$ in $E$}
        \State {Randomly choose $r \in_R [0,1]$ from a uniform dist.}
        \If{$r \leq w_{u,v}$}
            \State $E' \leftarrow E' \cup \{u,v\}$
        \EndIf
    \EndFor
  
    \State Construct $G' = (V, E')$
    \State \Return $G'$
\end{algorithmic}
\end{algorithm}

\begin{algorithm}
\caption{\sc{MixGreedy}($G,K,\mathcal{R}$)}
\label{alg:mixgreedy}
\algorithmicrequire{$G = (V,E)$: the influence graph
\\\hspace*{6.6ex}{$K$: number of seed vertices
\\\hspace*{6.7ex}$\mathcal{R}$: number of MC simulations per seed vertex}\\}
\algorithmicensure{$S$: a seed set that maximizes influence on $G$}
\begin{algorithmic}[1]
    \State $S, mg \leftarrow$ {\sc NewGreedy}$(G, 1, \mathcal{R})$
   \State {$\sigma_{G}(S) \leftarrow  \max_{v\in V}\{mg_v\}$} \algorithmiccomment{$mg_v = \sigma_{G}(\emptyset, v)$}
    \State $Q \leftarrow$ PriorityQueue()  
    \For{$v \in V\setminus S$}
        \State $Q$.enqueue($v$, priority=$mg_v$) 
    \EndFor
      \State $iter_v \leftarrow 0, \forall v\in V $
    \While{$|S| < K$} \label{ln:mg:celf1}
        \State $u \leftarrow Q$.top()
        \If{$iter_u = |S|$}
            \State $S \leftarrow S \cup \{u\}$
            \State $Q$.dequeue($u$)
            \State $\sigma_{G}(S) \leftarrow \sigma_{G}(S) + mg_u$          
        \Else
            \State $mg_u \leftarrow$ {\sc RandCas}$(G, S \cup \{u\}, \mathcal{R})- \sigma_{G}(S)$ \label{ln:randcas}
            \State $iter_u \leftarrow |S|$
            \State $Q$.updatePriority($u$, priority=$mg_u$) \label{ln:mg:celf2}
        \EndIf
    \EndWhile
    \State \Return $S$
\end{algorithmic}
\end{algorithm}

The pseudocode of {\sc MixGreedy} is given in Algorithm~\ref{alg:mixgreedy}. Note that {\sc MixGreedy} uses only a single iteration of {\sc NewGreedy} with parameters ($G, 1, {\cal R}$). Even though {\sc NewGreedy} can be used to find each of the $K$ vertices in $S$ one by one, Chen~et~al.'s experiments revealed that {\sc NewGreedy} is only faster in the initialization stage. For consequent vertices, the experiments show that performing the CELF approach and adding a vertex to the seed set in case of a revisit in the queue is faster.   

\begin{algorithm}[!ht]
\caption{\sc{RandCas}($G$, $S$, $R$)}
\algorithmicrequire{$G = (V, E)$: the influence graph\\
\hspace*{7ex}$S$: the seed set}\\
\algorithmicensure{$\sigma_G(S)$: influence score of seed set $S$ on $G$}

\begin{algorithmic}[1]
  \State $\sigma_S \leftarrow 0$
  \For{$r=1\ldots \mathcal{R}$}
    \State $G' = (V, E') \leftarrow$ {\sc Sample}($G$)
    \State Compute $R_{G'}(S)$
    \State $\sigma_G(S) \leftarrow \sigma_G(S)  + \frac{|R_{G'}(S)|}{{\cal R}}$
    \EndFor
    \State \Return $\sigma_G(S)$

\end{algorithmic}
\end{algorithm}


In this work, we propose \acro{}, the fused and restructured form of {\sc MixGreedy}. The memory accesses and floating-point operations performed by the existing algorithm are restructured to reduce the memory pressure for the marginal gain computations. This enables fused-sampling and vectorization. Furthermore, memoization is applied to reduce the cost of the CELF phase. The proposed techniques in this paper can be adopted by other probabilistic graph algorithms, as well as other IM kernels, to boost their performance. Although they are not focusing on probabilistic algorithms and fusing, SIMD-based alterations of graph kernels to regularize memory accesses have been studied before, e.g., to compute centrality metrics~\cite{SariyuceSKC14, SariyuceSKC15}. 

\subsection{Single instruction multiple data (SIMD)} 
{\em Single Instruction-Multiple Data} architectures allow parallelism at the instruction level. Initially started with 128-bit MMX vector extensions, many enhancements have been implemented in modern processors. In this work, we employed Advanced Vector Extensions (AVX2) instruction set. AVX2 works on 256-bit registers in many packed forms including 1x256, 2x128, 4x64, and 8x32 storage patterns. We added these vector instructions manually to the code since, 
even though compilers translate and optimized most of the loops to vectorized forms, {\em compare and move-mask} operations were not recognized by auto-vectorization in our preliminary experiments. For completeness, the intrinsics explicitly used in this paper are described in Table~\ref{tab:avx2-instructions}.

\begin{table}[!ht]
    \caption{AVX2 intrinsics used in the implementation.}
    \label{tab:avx2-instructions}
    \centering

    \begin{tabular}{|p{0.33\linewidth}|p{0.58\linewidth}|}
    \hline
    Intrinsic & Definition\\
    \hline
    {\tt \_mm256\_set1\_epi32} & Initializes 256-bit vector with scalar integer values. Doesn't map to any AVX instructions. \\
    {\tt \_mm256\_and\_si256} & Performs bitwise logical AND operation on 256-bit integer vectors. \\
    {\tt \_mm256\_xor\_si256} &  Performs bitwise logical XOR operation on 256-bit integer vectors.\\
    {\tt \_mm256\_cmpgt\_epi32} & Compares packed 8x 32-bit integers of two input vectors. \\
    {\tt \_mm256\_movemask\_ps} & Extracts the first bits of 8x 32-bit elements in a compact 8-bit format \\
    {\tt \_mm256\_blendv\_epi8} & Blends/selects byte elements of input vectors depending on the bits in a given mask vector.\\ 
             \hline         
    \end{tabular}
\end{table}{}

\section{Boosting Influence Maximization}\label{sec:infuser}

Classical Monte-Carlo based IM algorithms first sample a sub-graph and then perform a single simulation. Such an approach is amenable to thread-level, coarse-grain parallelization since the simulations are independent of each other. However, this requires the graph to be read from the memory for every simulation. 
The state-of-the-art implementations use this {\em one-sample-per-simulation} approach and build a unique graph for every sample to find the marginal influence scores~\cite{MixGreedy}. With coarse-grain parallelization, this makes the IM kernels inefficient in terms of performance since the graphs are sparse~(and samples are sparser), memory accesses are irregular, and performing a single simulation per graph traversal increases the already hindering pressure on the memory subsystem and makes the IM process further memory bound. As mentioned before, to make the IM computations faster, heuristics, sketches, and proxy models have been proposed in the literature. Unlike these, \acro{} exploits the properties of the greedy Monte-Carlo algorithm. It is tuned for the undirected graphs and the Independent Cascade model. However, the techniques such as fusing can be adopted by the other models or Monte-Carlo graph algorithms using sampling. 
\acro leverages three techniques to achieve its goals. 

\begin{itemize}[leftmargin=*]
    \item Instead of explicitly constructing a data structure for each subgraph, the proposed approach uses direction-oblivious pseudo-random numbers throughout the edge-based simulation to fuse the sampling with the computation of influence scores.
    \item To reduce the memory subsystem pressure, \acro leverages batched simulations and instruction-level parallelism and when possible, utilizes each edge access for multiple simulations. 
    \item To reduce the number of operations performed, the component IDs for each vertex and sampled subgraph are memoized which can then be used while computing the marginal gains during the CELF stage.
\end{itemize}

On top of these, multi-core parallelism is applied to further increase the performance by running multiple threads and assigning each batch to a different thread. 


\subsection{Direction oblivious hash-based sampling}

Traditionally, the cascade model requires a new sample, i.e., a subgraph, from $G = (V, E)$ to simulate the diffusion process. State-of-the-art implementations sample edges from $E$ and add them to a set along with reversely oriented edges to make the subgraph, which is constructed from this sampled edge set, undirected. 
\acro does not explicitly sample. Whenever an edge with a certain orientation is read from the memory, it is sampled or skipped depending on the outcome of direction-oblivious sampling that assigns the same sampling probability for both directions, $(u, v)$ and $(v, u)$. We utilize a hash function $h(u, v) = h(v,u)$ to get the same probability for forward and backward directions within the same simulation. The hash function used is 
\begin{equation}
    \label{eq:hash}
    h(u,v) = \mbox{{\sc Murmur3}}(\min(u,v) || \max(u,v))  
\end{equation}
where $||$ is the concatenation operator. To avoid the cost of hashing during simulations, all possible hash values are pre-computed. Although there exist $\frac{n \times (n-1)}{2}$ possible vertex pairs, we only need the vertex pairs having an edge in between, i.e., only $m$ hash values are pre-computed. We have tried a few other hash algorithms as well; we chose {{\sc Murmur3}}~\cite{MurmurHash3} due to its simplicity and good avalanche behavior with maximum bias $0.5\%$.

Although the above-mentioned approach generates a unique hash value for each edge, and hence a unique sampling probability, different simulations require different probabilities. To achieve this, we use a random number $X_r$ for each simulation $r$. To compute the sampling probability of $\{u, v\}$ during $r$th simulation, $h(u,v)$ is first XOR'ed with a uniformly randomly chosen $X_r \in_R [0, h_{max}]$ and the outcome is divided to the maximum possible hash value $h_{max}$. Let ${\rho}(u,v)_r$ denote this sampling probability for $\{u,v\}$ in simulation $r$. Formally,
\begin{equation}
    \label{eq:hash_prob}
    {\rho}(u,v)_r = \frac{X_r \oplus h(u,v)}{h_{max}}.
\end{equation}
The edge $\{u,v\}$ is verified to be in the sample if ${\rho}(u,v)_r$ is smaller than or equal to the threshold $w_{u,v}$. With the proposed approach, sampling an edge reduces to an XOR and compare-greater-than operation. The branching on the latter can be removed to enable SIMD instructions as explained later in this section. 

\begin{figure}[!ht] 
    \centering
    \includegraphics[width=1\linewidth]{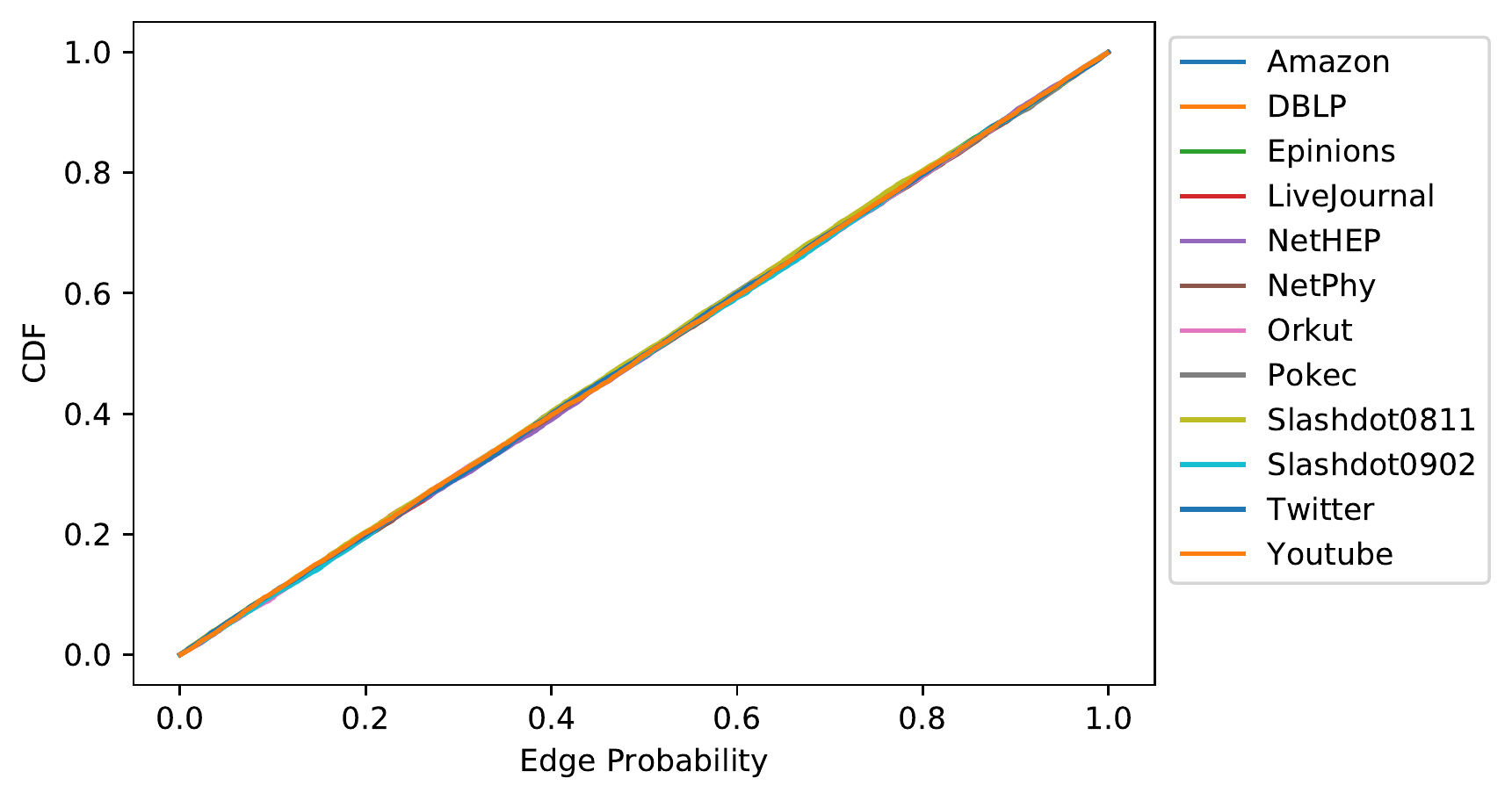}
    \caption{Cumulative distribution function of hash-based sampling probabilities on various real-life networks.}
    \label{fig:prob_cdf} 
\end{figure}

 {{\sc Murmur3}} guarantees a change on the $50\%$ of the bits when a single bit of the input changes. Furthermore, all bits independently change when the input is changed. These properties allow us to generate good pseudo-random values to simulate the process. For practical considerations, we stored all the $\rho(u,v)_r$s generated for various real-life networks and plotted the Cumulative Distribution Function~(CDF) of these values. For a given graph $G = (V,E)$, the CDF of a sampling probability $x$ is computed as $\Pr\left(x \leq \rho(u,v)_r\right)$ for all $\{u,v\} \in E$ and $0 \leq r < R$. Figure~\ref{fig:prob_cdf} shows the CDFs for 12 real-life networks. The sampling probability distribution with hash-based computation is almost identical with the uniform distribution which is required to simulate the diffusion process.

In \acro{}, the diffusion is performed on a subgraph which is never constructed; in fact, each diffusion is simulated on $G$. Thus the overhead of generating and storing a sample and reading it back from the memory is avoided. However, for each visit of $\{u, v\}$, since \acro{} does not know if the edge is in the sample or not, ${\rho}(u,v)_r$ is recomputed. Another immediate benefit of fusing is traversing only the vertices that contribute to influence score and their neighbors. On the other hand,  a non-fused implementation would traverse all edges for all simulations. Often, the total influence is a very small fraction of the total number of vertices and hence, fusing is vital to have a scalable IM kernel.


\subsection{Vectorized Monte-Carlo graph traversal}

\begin{figure}[!ht] 
    \centering
    \subfloat[\label{fig:sims}]{%
        \includegraphics[width=0.47\linewidth]{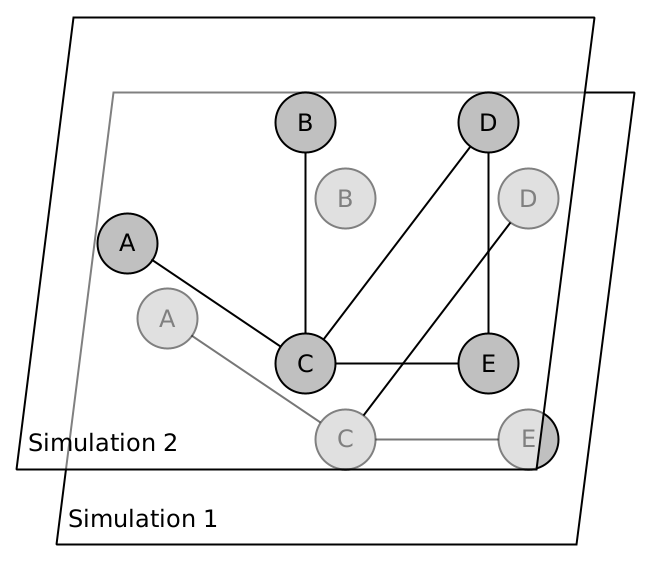}
    } 
    \subfloat[\label{fig:fused}]{%
     \includegraphics[width=0.50\linewidth]{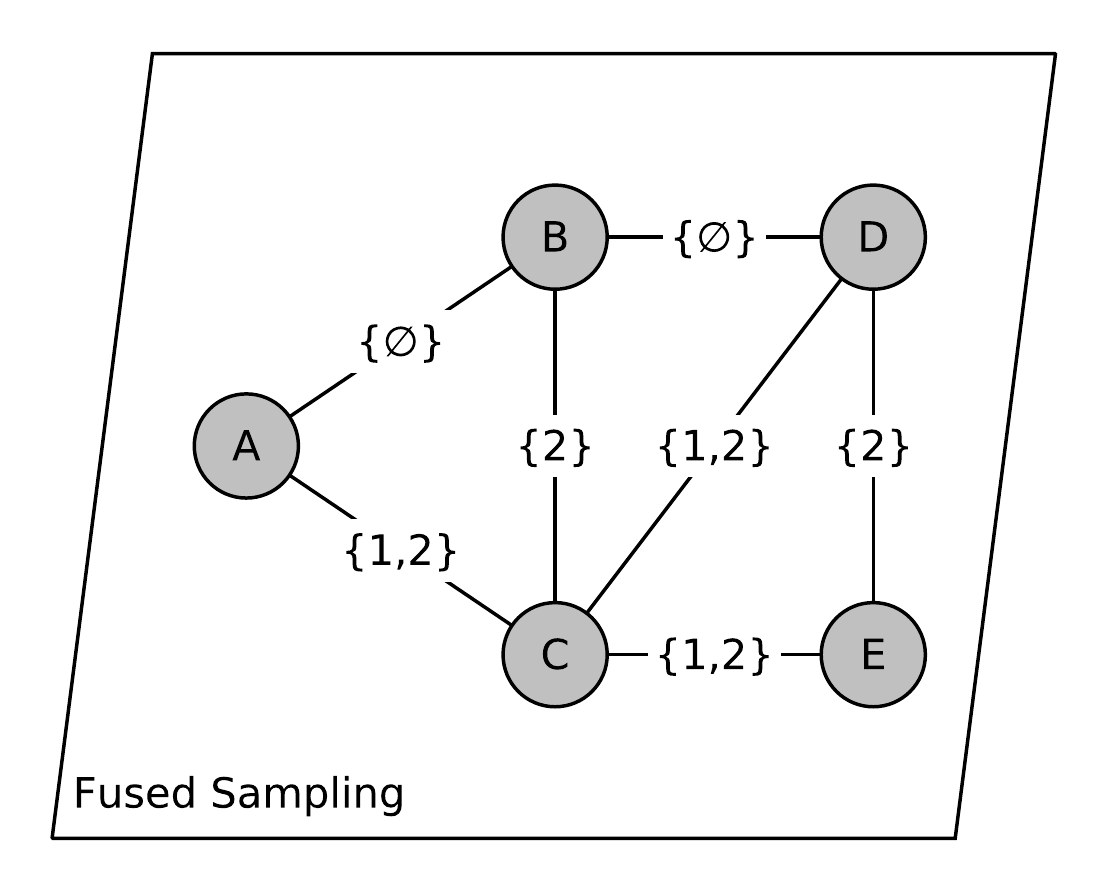}}
  \caption{
  \protect\subref{fig:sims} Two sampled subgraphs of the toy graph from Figure~\ref{fig:ic} with 5 vertices and 10 edges.
  \protect\subref{fig:fused} The simulations are performed in a way to be fused with sampling. Each edge is labeled with the corresponding sample/simulation IDs. 
  }
  \label{fig:traversal} 
\end{figure}

In {\sc MixGreedy}~(Algorithm~\ref{alg:mixgreedy}), both the {\sc NewGreedy} step and marginal gain computations utilize graph sampling. 
By leveraging vectorization, a single thread in \acro{} can process a batch of $B$ samples/simulations at once. A high-level visualization of how the samples are batched is given in Figure~\ref{fig:traversal}.
In a perfect, fused, and batched execution, the edges (of the original graph) flow from the memory to the cores and they are consumed by carefully structured SIMD kernels. Once an edge is visited, all ${\cal R}$ simulations are taken into account by batches of $B$ simulations. Although fusing and vectorization can incur redundant computations, as the experiments will show, the proposed approach significantly boosts the performance. 

\subsubsection{A vectorized {\sc NewGreedy} step}

For an undirected graph, the {\sc NewGreedy} step of {\sc MixGreedy} needs to identify the reachability sets $R_{G'}(v)$ for all $v \in G'$. Traditional IM implementations work on a single subgraph and initiate many graph traversals until all vertices are visited. The time complexity of this process is linear in terms of the number of vertices and edges. However, its memory access pattern tends to be irregular; many random memory accesses are required which results in low CPU utilization. 
Instead of graph traversal, e.g., Breadth-First Search, the connected components within a sampled subgraph can be found via {\em label propagation}, which starts by assigning unique labels to each vertex. Then at each iteration, the edges are visited and the labels of both endpoints are set to the minimum of the two. This process continues until convergence; i.e., no label is changed within a single iteration. The total amount of work performed by this algorithm is superlinear since each edge is touched at each iteration. To reduce the time complexity, one can mark the ({\em live}) vertices whose labels are updated in the current step, and only process their edges in the next step. Although this does not guarantee a linear-time algorithm, it significantly reduces the number of edge accesses.  

\begin{figure}[!ht]
\includegraphics[width=\linewidth]{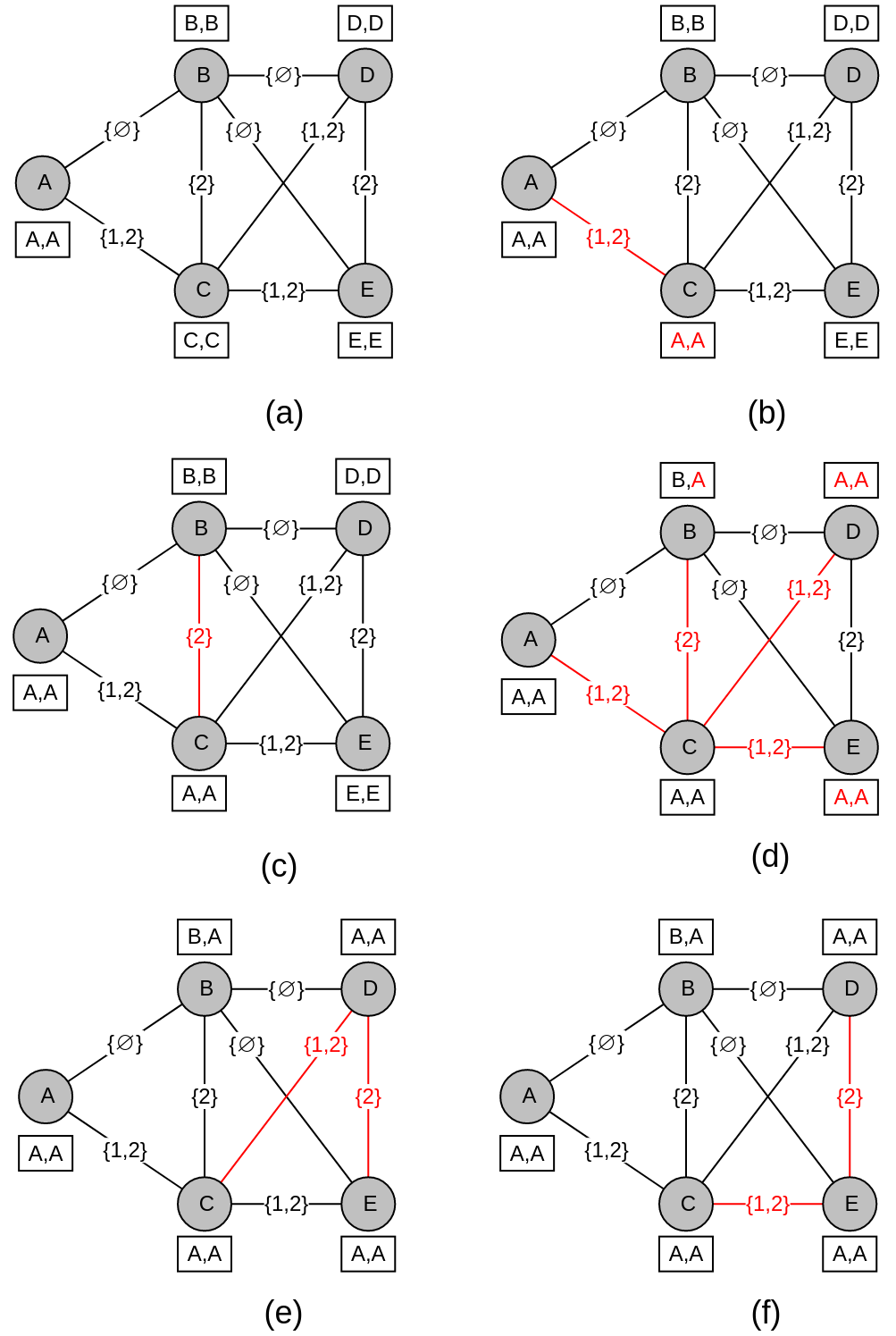}
\caption{(a) The initial state on a toy graph for label propagation; all vertices are labeled with their ids. (b) First, the edges of $A$ are processed; the edge to $C$ is in both samples. $C$'s labels are updated. (c) $B$'s edges are processed. The edge to $C$ exists in the second sample. $C$'s second label is smaller, hence no update is performed. (d) $C$'s edges are being processed. It has edges to $A$, $B$, $D$, and $E$ in the samples. The labels <$A$, $A$> are propagated to $D$ and $E$ since the edges are in both samples. Besides, $B$'s second label is updated because only sample 2 contains the corresponding ($C$, $B$) edge. (e)(f) $D$ and $E$ edges in the samples. However, the are no updates.} \label{fig:ng-processing}
\end{figure}

\acro{} runs the above-mentioned, label-propagation-based approach in a fused and batched manner. For all ${\cal R}$ samples, the propagation is simulated on the original graph $G$ by taking only the sampled edges into account. The simulations are processed on batches of $B = 8$ samples which are never constructed. To do that, the existence of the edge in these samples is rechecked every time it is being processed. All the live vertices within a single iteration are processed in parallel by multiple threads. Further parallelization at this stage comes from running $B$ simulations at once in a SIMD fashion. An example run with ${\cal R} = B = 2$ simulations is given in  Figure~\ref{fig:ng-processing} continuing from Figure~\ref{fig:traversal}.

\begin{algorithm}
\caption{\sc{NewGreedyStep-Vec}($G,\mathcal{R}$)}
\label{alg:vecnewgreedy}
\algorithmicrequire{$G = (V,E)$: the influence graph\\
\hspace*{6.8ex}$\mathcal{R}$: number of MC simulations per seed vertex}\\
\algorithmicensure{ $mg$: marginal influence scores \\
\hspace*{8.8ex} $l$: connected component labels}

\begin{algorithmic}[1]
 \For{$v\in V$}\label{ln:lab1}
        \State $l_v \leftarrow [v,\ldots, v]_{\cal R} $\label{ln:lab2}
        \EndFor
            \State{${\cal L} \leftarrow V$}
		    \While {${\cal L}$ is not empty}
		      \State{${\cal L'} \leftarrow \emptyset$}

		        \For{ $u \in {\cal L}$ {\bf in parallel}} \label{ln:for1}
                    \For{ $v \in \Gamma_G(u)$ } \label{ln:for2}
                        \State $r \leftarrow 0$
                \While{$ r < \mathcal{R}$}
                        \For{$r' = r \ldots r + 7$}\label{ln:startngv} \Comment{$B = 8$}
                            \If{$\rho(u,v)_{r'} \geq w_{u,v}$}
                            \State $min_{label} \leftarrow {\tt min}(l_{u}[r'],l_{v}[r'])$ 
                                \If{$min_{l} \neq l_{v}[r']$} 
                                     \State{$l_{v}[r'] \leftarrow min_{l}$}\label{ln:update}
                                     \State{${\cal L}' \leftarrow {\cal L}' \cup \{v\}$}\label{ln:endngv}
                                \EndIf
                            \EndIf
                        \EndFor
                    \State $r \leftarrow r + 8$
                \EndWhile

                    \EndFor
                \EndFor
                \State${\cal L} \leftarrow  {\cal L}'$

		    \EndWhile
        \For {$v \in V$ {\bf in parallel}}
                \State $mg_v \leftarrow 0 $
        \For {$r = 1 \ldots {\cal R}$}
                \State $mg_{v} \leftarrow  mg_{v} + |\{u: l_{u}[r] = l_{v}[r]\}|$	         
	       \EndFor

         \EndFor
    \State \Return $mg$, $l$
\end{algorithmic}
\end{algorithm}

 \begin{algorithm}
\caption{\sc{VecLabel} ({$r, u,v,X_r,l_u,l_v,r$})}\label{alg:VecLabel}
\algorithmicrequire{
$r$: ID of the first simulation in the current batch\\
\hspace*{7ex}$u:$ source vertex\\
\hspace*{7ex}$v:$ target vertex\\
\hspace*{7ex}$X_r$: random number for simulations $r$ to $r + 7$\\
\hspace*{7ex}$l_u$: vector of component labels of $u$\\
\hspace*{7ex}$l_v$: vector of component labels of $v$}\\
\algorithmicensure{$l_v$: labels of vertex $v$ after traversing edge $(u,v)$}\\
\hspace*{8ex}$live_v$: a boolean which state if $v$ is live

\begin{algorithmic}[1]

    \State $mask \leftarrow {\tt \_mm256\_cmpgt\_epi32}(l_u[r],l_v[r])$
	\State $labels \leftarrow {\tt\_mm256\_blendv\_epi8}(l_u[r],l_v[r],mask)$
	\State $hashes \leftarrow {\tt\_mm256\_set1\_epi32}(hash(u,v))$
	\State $probs \leftarrow {\tt\_mm256\_xor\_si256}(hashes,\ X_r)$
 	\State $w_{vec} \leftarrow {\tt\_mm256\_set1\_epi32}(\lfloor w_{u,v} \times {\tt INT\_MAX}\rfloor)$
	\State $select \leftarrow {\tt\_mm256\_cmpgt\_epi32}(w_{vec},\ probs)$
    \State $l_v[r] \leftarrow {\tt\_mm256\_blendv\_epi8}(l_v[r],\ labels,\ select)$
    \State $live_v \leftarrow {\tt\_mm256\_movemask\_ps}(\newline
        \hspace*{4em}{\tt \_mm256\_and\_si256}(select, mask))$
    \State \Return $l_v$, $live_v$
\end{algorithmic}
\end{algorithm}


Algorithm~\ref{alg:vecnewgreedy} describes the fused and vectorized {\sc NewGreedyStep-Vec}. The algorithm takes two inputs $G$, the original graph, and ${\mathcal R}$, the number of simulations. It works along the same lines with the original {\sc NewGreedy} with additional operations for label propagation. The labels for each vertex are initially set as the vertex IDs~(lines~\ref{ln:lab1}--~\ref{ln:lab2}). The outer {\bf while} loop checks if there exist any more live vertices. Here, a vertex is said to be {\em live} if at least one of its ${\cal R}$ labels is changed during the previous iteration. The first inner {\bf for} at line~\ref{ln:for1} loops over the live vertices in a multi-threaded fashion. A single thread runs the next {\bf for} loop at line~\ref{ln:for2} to visit the edges of the live vertex being processed. The operations corresponding to each of the ${\cal R}$ simulations are performed for a visited edge $(u,v)$ in batches of 8. 

For each $0 \leq r < {\cal R}$, where $r$ is a multiple of $B = 8$, the vectorized steps that perform the operations in simulations $r$ to $r + 7$ are given between lines~\ref{ln:startngv}--\ref{ln:endngv}. These steps are performed as described in Algorithm~\ref{alg:VecLabel}, {\sc VecLabel}. The algorithm first compares the labels using element-wise compare intrinsic ${\tt \_mm256\_cmpgt\_epi32}$ which returns all 1's~($2^{32}-1$) when the first value is larger, and 0 otherwise.
Then, pairwise minimum of the labels from the two vectors can be selected by ${\tt \_mm256\_blendv\_epi8}$ that employs the $mask$ entries generated by the previous step. This intrinsic selects the bytes from the first vector if the corresponding $mask$ entry is not zero. Otherwise, it selects the bytes from the second vector. Hence for an edge $(u, v) \in E$, the resulting vector, $labels$, contains the smaller of the endpoints', i.e., $u$'s and $v$'s labels, for each simulation. The edge $(u, v)$ may not have been sampled by all simulations. To find the simulations it is sampled, the algorithm generates the sampling probabilities by XORing the corresponding hash, $h(u,v)$, and the random values, $X_r$. 
Being computed in the preprocessing step, the hash is promoted to a vector, $hashes$, by the intrinsic ${\tt \_mm256\_set1\_epi32}$. The XOR operations are performed in a SIMD fashion with the intrinsic ${\tt \_mm256\_xor\_si256}$. We then promote $w_{u,v}$ to a vector $w_{vec}$ by first multiplying it with {\tt INT\_MAX} using the ${\tt \_mm256\_set1\_epi32}$ intrinsic. Then, this vector is element-wise  compared to the vector $probs$ by using ${\tt \_mm256\_cmpgt\_epi32}$. The result of this operation is the $select$ vector containing the selection masks for simulations. Blending $labels$~(from line 2) with $v$'s current labels based on the $select$ entries produces $v$'s final labels for the corresponding simulations $r$ to $r + 7$ by using the intrinsic ${\tt \_mm256\_blendv\_epi8}$.

After the new labels are computed, we check if any of $v$'s labels are modified to verify whether the process is converged or not. To do this, we first perform bitwise-and operations for the elements in $mask$ and $select$ by using the intrinsic ${\tt \_mm256\_and\_si256}$. Then, the first bits of the 32-bit elements are extracted in a compact 8-bit format by using the ${\tt \_mm256\_movemask\_epi8}$ intrinsic. This intrinsic eliminates 8 comparison branches and produces a {\bf boolean} variable $live_v$. As mentioned above, at each iteration, the algorithm only processes the vertices whose labels are changed in the previous iteration. Initially, all the vertices are considered live. Each thread uses these $live_v$ values to keep track of the set $\mathcal{L}$ of live vertices. To do this, we use an array of size $n$ in which the $v$th entry is marked if $v$ is live. After an iteration is finished, $\mathcal{L}$ is updated. This approach allows us only to process live vertices.

\subsection{Finding marginal gains with memoization}


During the label propagation stage in {\sc NewGreedyStep-Vec}, \acro{} computes and stores all the component labels $l$~(obtained by concatenating each $l_v$ for all $v \in V$) that can be considered as a two-dimensional $n  \times {\cal R}$ array. The first seed vertex is indeed the one having the largest expected~(average) component size. Instead of resampling, this information can be utilized during the CELF stage while computing marginal gains and finding the remaining $K-1$ seed vertices. The marginal gain for a vertex $u$, i.e., $mg_u$, can be found by computing the average number of vertices~(over all the ${\cal R}$ samples) that belong to $u$'s connected component but do not belong to the components of the seed vertices. This is equal to the expected number of additional vertices that will be influenced by inserting $u$ to the seed set $S$. 

While computing $mg_u$, for all simulations, one can compare $u$'s label to all the component labels of the seed vertices in respective simulation. In our implementation, the data structure $l$ is stored as a single large memory block where the ${\cal R}$ labels of a single vertex are stored consecutively for a better spatial locality. The component sizes are also stored in similar two-dimensional $n \times {\cal R}$ array where rows correspond to component labels and columns correspond to simulations. Labels that do not map to a component are wasted for fast access while keeping the asymptotic space complexity the same (as $l$'s space complexity). This process is equivalent to using {\sc RandCas} over existing ${\cal R}$ samples for finding marginal gains, except, no graph traversal or sampling is performed. Compared to the original approach, the memory accesses are more regular and the cache is better utilized. Furthermore, this operation can be efficiently parallelized as shown in the pseudocode of \acro{}, Algorithm~\ref{algo:infusermg}~(lines~\ref{ln:mgv:1}--\ref{ln:mgv:2}).

\begin{algorithm}
\caption{\sc{\acro}($G,K,\mathcal{R}$)}
\label{algo:infusermg}
\algorithmicrequire{$G$: Graph \\ 
\hspace*{7ex}$K$: size of the seed set\\
\hspace*{7ex}$\mathcal{R}$: number of MC simulations to perform}\\
\algorithmicensure{ $S$: a seed set that maximizes influence}

\begin{algorithmic}[1]
    \State $mg,l \leftarrow ${\sc{ NewGreedyStep-Vec}}$(G, \mathcal{R})$
    \State $S \leftarrow \{\varnothing \}$
    \State $q \leftarrow$ PriorityQueue()  
    \State $iter_v \leftarrow 0, \forall v\in V $
    \For{$v \in V$}
        \State $q.enqueue$($v$, priority=$mg_v$) 
    \EndFor
    \State $R_{G'}(v) \leftarrow 0, \forall v\in V $
    \While{$|S| < K$}
        \State $u \leftarrow q.dequeue()$
        \If{$iter_u = |S|$}
            \State $R_{G'}(S) \leftarrow R_{G'}(S \cup \{u\})$ \Comment{Append $l_u$ to $R_{G'}(S)$}
            \State $S \leftarrow S \cup \{u\}$  \Comment{Commit $u$ into $S$}
        \Else
            \State $mg_u \leftarrow 0$
            \For{$r = 1$ to ${\cal R}$ {\bf in parallel reduce($mg_u$)}} \label{ln:mgv:1}
            \State $mg_u \leftarrow mg_u + |l_{u}[r] \in \{l
            \setminus R_{G'_r}(S)\}|$ \label{ln:mgv:2}
            \EndFor
            \State $iter_u \leftarrow |S|$
            \State $q.enqueue$($u$, priority=$mg_u$)
        \EndIf
    \EndWhile
    \State \Return $S$

\end{algorithmic}
\end{algorithm}


\subsection{Implementation Details}
All the algorithms use the Compressed Sparse Row~(CSR) graph data structure. In CSR, an array, $xadj$, holds the starting indices of each vertices neighbors, other vector, $adj$, holds neighbors of each vertex consecutively. So, to reach neighbors of vertex i, first we visit $xadj[i]$ and $index[i+1]$ to find start and end positions in data, then scan starting from $adj[index[i]]$ until $adj[index[i+1]]$ position. 


\section{Experimental Results} \label{sec:exp}
All the experiments are performed on a server equipped with two 8-core Intel Xeon CPU E5-2620v4 sockets running on 2.10GHz and 192GB memory. Hence, there exist 16 cores in total. The OS running on the server is {\tt Ubuntu 16.04.2 LTS} with {\tt Linux 4.4.0-66} generic kernel. The algorithms are implemented in {\tt C++} and compiled with {\tt gcc 8.2.0} with {\tt -Ofast} as the optimization flag. Multi-threaded CPU parallelization is obtained with {\tt OpenMP} pragmas. We have manually utilized AVX2 instructions available on the CPUs by using compiler intrinsics to implement the algorithms.

\begin{table}[!ht]
\caption{Properties of networks used in the experiments}\label{tab:NetProps}
\centering
\scalebox{0.90}{
\begin{tabular}{ll||r|r|r|r}
& & No. of           & No. of    & Avg.  &  Avg.  \\
&Dataset & Vertices          & Edges     &    Weight         &  Degree               \\
\hline
\multirow{6}{*}{\rotatebox[origin=c]{90}{Undirected}}& {\tt Amazon} & 262,113 & 1,234,878 & 1.00 & 4.71 \\
&{\tt DBLP} & 317,081 & 1,049,867 & 1.00 & 3.31 \\
&{\tt NetHEP} & 15,235 & 58,892 & 1.83 & 3.87 \\
&{\tt NetPhy} & 37,151 & 231,508 & 1.28 & 6.23 \\
&{\tt Orkut} & 3,072,441 & 117,185,083 & 1.00 & 38.14 \\
&{\tt Youtube} & 1,134,891 & 2,987,625 & 1.00 & 2.63 \\

\hline
\multirow{6}{*}{\rotatebox[origin=c]{90}{Directed}}&{\tt Epinions} & 75,880 & 508,838 & 1.00 & 6.71 \\
&{\tt LiveJournal} & 4,847,571 & 68,993,773 & 1.00 & 14.23 \\
&{\tt Pokec} & 1,632,803 & 30,622,564 & 1.00 & 18.75 \\
&{\tt Slashdot0811} & 77,360 & 905,468 & 1.00 & 11.70 \\
&{\tt Slashdot0902} & 82,168 & 948,464 & 1.00 & 11.54 \\
&{\tt Twitter} & 81,306 & 2,420,766 & 1.37 & 29.77 
\end{tabular}
}
\end{table}

\begin{table*} 
\caption{Execution times~(in secs), memory use~(in GBs), and influence scores of the algorithms on the networks with $K = 50$ seeds and constant edge weights with $p = 0.01$.}
\label{tab:timings}
\centering
\scalebox{0.90}{
\begin{tabular}{l||r|r|r||r||r|r|r||r|r|r}
  & \multicolumn{4}{c||}{Execution time in seconds.} & \multicolumn{3}{c||}{Memory use in Gigabytes.}& \multicolumn{3}{c}{Influence scores.}\\
  & \multicolumn{1}{c|}{{\sc Mix}} &\multicolumn{1}{c|}{{\sc Fused}}   &   \multicolumn{1}{c||}{\sc INFuser} & \multicolumn{1}{c||}{\sc INFuser} & \multicolumn{1}{c|}{{\sc Mix}} &\multicolumn{1}{c|}{{\sc Fused}}   &   \multicolumn{1}{c||}{\sc INFuser} & \multicolumn{1}{c|}{{\sc Mix}} &\multicolumn{1}{c|}{{\sc Fused}}   &   \multicolumn{1}{c}{\sc INFuser} \\
 & \multicolumn{1}{c|}{{\sc Greedy}} &\multicolumn{1}{c|}{{\sc Sampling}}  &  \multicolumn{1}{c||}{\sc MG}&  \multicolumn{1}{c||}{($K = 1$)} & \multicolumn{1}{c|}{{\sc Greedy}} &\multicolumn{1}{c|}{{\sc Sampling}} &  \multicolumn{1}{c||}{\sc MG}& \multicolumn{1}{c|}{{\sc Greedy}} &\multicolumn{1}{c|}{{\sc Sampling}} &  \multicolumn{1}{c}{\sc MG}\\
Dataset  & \multicolumn{1}{c|}{($\tau=1$)} & \multicolumn{1}{c|}{($\tau=1$)} & \multicolumn{1}{c||}{($\tau=16$)} & \multicolumn{1}{c||}{($\tau=16$)}& \multicolumn{1}{c|}{($\tau=1$)} & \multicolumn{1}{c|}{($\tau=1$)}  & \multicolumn{1}{c||}{($\tau=16$)}& \multicolumn{1}{c|}{($\tau=1$)} & \multicolumn{1}{c|}{($\tau=1$)}  & \multicolumn{1}{c}{($\tau=16$)}  \\
\hline
{\tt Amazon}       &    141.31 & 48.84    & 2.09  & 2.09 &     0.76 &  0.18 &         4.05  & 158.28    &  158.63    &  158.63 \\
{\tt DBLP}         &         - & 305.38   & 7.02  &  7.00   &       - &  0.25&      6.56    & - & 245.43   & 245.43  \\
{\tt Epinions}     &         - & 157069.53& 1.91  &   1.53  &      - &  0.05&       1.18    & - & 3051.39    &  3051.39 \\
{\tt LiveJournal}  &         - & -        & 265.84&  218.28 &       - &  -&  75.38    & - &  -   &  260364.56 \\
{\tt NetHEP}       &    259.05 & 12.60    & 0.08  &  0.07   &     2.27  &    0.01 & 0.24  &  132.38   & 136.45 & 136.45  \\
{\tt NetPhy}       &   1725.15 &  247.21  & 0.36  &  0.34 &     8.56 &  0.03 &    0.58 &  312.56    & 332.52 & 332.52\\
{\tt Orkut}        &         - &  -       & 654.52&  586.55 &       - &  -&  50.45    & - &  -  & 650237.06 \\
{\tt Pokec}        &         - &  -       & 227.24& 196.85 &        - &  -&  26.02    & - & -   &  104196.34 \\
{\tt Slashdot0811} &         - & 211783.43& 2.69  &  2.00 &       - &    0.07&   1.21 & -     &  5197.88 & 5197.88\\
{\tt Slashdot0902} &         - & 233822.30& 3.11  & 2.17 &    - &   0.08&    1.29  &  -   & 5432.14 & 5432.14 \\
{\tt Twitter}      &         - &  -       & 3.07  & 2.11 &        - &  -&   1.32  &  -    & -  & 12441.56\\
{\tt Youtube}      &         - &  -       & 26.18 &  20.85&       - &   -&  17.83   &  -  &  -    &   9139.01 
\end{tabular}
}
\end{table*}

\begin{table*} 
\setlength{\tabcolsep}{3pt}
\caption{Execution times~(in secs) of the algorithms with $K = 50$ seeds in different simulation settings.}
\label{tab:timings-speedup}
\centering
\scalebox{0.90}{
\begin{tabular}{l||r|r|r||r|r|r||r|r|r||r|r|r}
  & \multicolumn{3}{c||}{$p = 0.01$} & \multicolumn{3}{c||}{$p = 0.1$} & \multicolumn{3}{c||}{$p \in N(0.05,0.025)$} & \multicolumn{3}{c}{$p \in [0,0.01]$}\\
  & \multicolumn{1}{c|}{{\sc Imm}} & \multicolumn{1}{c|}{{\sc Imm}}  & \multicolumn{1}{c||}{{\sc InFuseR}} & \multicolumn{1}{c|}{{\sc Imm}} & \multicolumn{1}{c|}{{\sc Imm}}  & \multicolumn{1}{c||}{{\sc InFuseR}} & \multicolumn{1}{c|}{{\sc Imm}} & \multicolumn{1}{c|}{{\sc Imm}}  & \multicolumn{1}{c||}{{\sc InFuseR}} & \multicolumn{1}{c|}{{\sc Imm}} & \multicolumn{1}{c|}{{\sc Imm}}  & \multicolumn{1}{c}{{\sc InFuseR}} \\
  Dataset & \multicolumn{1}{c|}{($\epsilon = 0.13$)} & \multicolumn{1}{c|}{($\epsilon = 0.5$)} & \multicolumn{1}{c||}{{\sc MG}}  & \multicolumn{1}{c|}{($\epsilon = 0.13$)} & \multicolumn{1}{c|}{($\epsilon = 0.5$)} & \multicolumn{1}{c||}{{\sc MG}}  & \multicolumn{1}{c|}{($\epsilon = 0.13$)} & \multicolumn{1}{c|}{($\epsilon = 0.5$)} & \multicolumn{1}{c||}{{\sc MG}} & \multicolumn{1}{c|}{($\epsilon = 0.13$)} & \multicolumn{1}{c|}{($\epsilon = 0.5$)} & \multicolumn{1}{c}{{\sc MG}} \\
    \hline
{\tt Amazon} & 62.67 & 4.95 & 2.09 & 24.80 & 2.72 & 9.99 & 8.64 & 0.84 & 3.24 & 8.15 & 1.29 & 3.56 \\ 
{\tt DBLP} & 55.92 & 4.02 & 7.02 & 168.68 & 15.34 & 11.83 & 46.90 & 4.97 & 11.28 & 56.34 & 5.02 & 12.66 \\ 
{\tt Epinions} & 72.39 & 7.55 & 1.91 & 86.10 & 7.82 & 1.96 & 92.28 & 9.58 & 1.29 & 91.68 & 9.08 & 1.10 \\ 
{\tt LiveJournal} & 9078.34 & 860.38 & 265.84 & - & 1527.58 & 153.46 & - & 1678.81 & 190.90 & - & 1732.58 & 214.98 \\ 
{\tt NetHEP} & 2.80 & 0.29 & 0.08 & 6.31 & 0.65 & 0.18 & 4.33 & 0.43 & 0.15 & 4.41 & 0.42 & 0.19 \\ 
{\tt NetPhy} & 3.55 & 0.39 & 0.36 & 22.57 & 2.06 & 0.73 & 18.07 & 1.64 & 0.79 & 16.55 & 1.69 & 0.69 \\ 
{\tt Slashdot0811} & 135.54 & 12.33 & 2.69 & 146.09 & 14.48 & 2.04 & 166.84 & 16.08 & 1.58 & 160.03 & 17.18 & 1.57 \\ 
{\tt Slashdot0902} & 107.83 & 10.63 & 3.11 & 129.15 & 13.29 & 1.81 & 151.31 & 13.59 & 1.97 & 145.54 & 14.74 & 1.56 \\ 
{\tt Orkut} & 24300.59 & 2279.10 & 654.52 & - & 1987.11 & 195.60 & - & 2541.79 & 270.43 & - & 2642.77 & 225.02 \\ 
{\tt Pokec} & 2646.98 & 247.36 & 227.24 & - & 611.36 & 74.38 & 8060.71 & 796.88 & 108.20 & 8477.91 & 735.35 & 96.11 \\ 
{\tt Twitter} & 298.97 & 26.70 & 3.07 & 261.94 & 23.70 & 2.52 & 321.48 & 30.10 & 1.85 & 310.16 & 30.44 & 1.91 \\ 
{\tt Youtube} & 201.65 & 19.42 & 26.18 & 740.35 & 78.51 & 26.31 & 643.09 & 61.08 & 32.45 & 649.30 & 61.80 & 25.18
\end{tabular}
}
\end{table*}

\subsection{Network datasets used in the experiments\label{sec:graphs}}

The experiments are performed on twelve graphs~(six undirected, six directed) that have been frequently used for Influence Maximization. For directed datasets, the reverse edges are added to obtain undirected variants. The datasets are {\tt Amazon} co-purchase network~\cite{snapnets}, {\tt DBLP} co-laboration network~\cite{snapnets}, {\tt Epinions} consumer review trust network, {\tt LiveJournal}~\cite{snapnets}, {\tt NetHEP} citation network~\cite{MixGreedy}, {\tt NetPhy} citation network~\cite{MixGreedy}, {\tt Orkut}~\cite{snapnets}, {\tt Pokec} Slovakian poker game site friend network~\cite{snapnets}, {\tt Slashdot} friend-foe networks~(08-11, 09-11)~\cite{snapnets}, {\tt Twitter} list co-occurence network~\cite{snapnets}, and {\tt Youtube} friendship network~\cite{snapnets}. The properties of these datasets are given in Table~\ref{tab:NetProps}. 

For a thorough experimental evaluation, four influence settings are simulated; for each network, we use 
\begin{enumerate}
    \item constant edge weights $p = 0.01$~(as in~\cite{kempe2003maximizing} and~\cite{MixGreedy}),
    \item constant edge weights $p = 0.1$~(as in~\cite{kempe2003maximizing}),
    \item uniformly distributed weights from the interval [0, 0.1],
    \item normally distributed weights with mean 0.05 and std. deviation 0.025 so that 95\% of the weights lie in [0, 0.1].
\end{enumerate}

\subsection{Metrics used to evaluate the performance}

Following the literature, we employ three metrics to evaluate an algorithm; (i) the influence score, i.e., the expected number of vertices that are influenced (ii) the execution time, (iii) maximum memory size. There is an interplay among these metrics; it is trivial to devise an ultra-fast IM algorithm with a bad influence score. Similarly, using more memory can make an algorithm avoid computations. We present these metrics for each algorithm on all graphs. 

When the algorithms run on the same machine, the reported execution times and memory usages of different algorithms are comparable. However, the reported influence scores can be misleading since the algorithms may be using different approaches to estimate the influence score. To find the expected number of vertices, we requested and used the original implementation from Chen~et~al.~\cite{MixGreedy} as an oracle with minor modifications; i.e., without logging and using heap memory instead of stack memory to handle large-scale graphs. The random values in the oracle are generated by {\tt C++}'s Mersenne Twister 32-bit pseudo-random generator {\tt mt19937}, with a state size of 19937 bits.

\subsection{Algorithms evaluated in the experiments}
 
The algorithms that are evaluated can be classified into three groups. The first class contains {\sc MixGreedy}, obtained from Chen~et~al.~\cite{MixGreedy}, which is also used as the oracle to compute the influence scores. The second class contains two variants from the current state-of-the-art, Minutoli et al.'s {\sc Imm}~\cite{minutoli2019fast}. {\sc Imm} is a fast algorithm robustly producing high-quality seed sets which can influence a large number of vertices. In the original paper, the variant with $\epsilon = 0.13$, a user-defined hyper-parameter controlling the approximation boundaries, is suggested. We use this variant along with a much faster one with $\epsilon = 0.5$, which is also experimented in~\cite{minutoli2019fast}. 

The third class of algorithms contains two \acro{} variants. To show the speedup breakdown, we consider each variant as a separate algorithm. The first variant is {\sc FusedSampling} which only integrates the sampling step by generating probabilities on the fly without any algorithmic improvements or edge traversal savings. This variant performs the simulations one-by-one as in {\sc MixGreedy}. The second variant is the proposed approach \acro{} employing vectorization and memoization. Both of these variants employ CELF and use the queue-based vertex processing as the base algorithm {\sc MixGreedy}.
 
In this section, we first compare the \acro{} variants with {\sc MixGreedy} to present the speedups over the baseline with fusing and vectorization. We then compare \acro{} with the state-of-the-art to better position the proposed approach in the literature. Last, we evaluate the multi-threading performance of \acro{} with $\tau \in \{1, 2, 4, 8, 16\}$ threads. In all experiments, we use a time-limit of 302,400 seconds~(3.5 days).

\subsection{Comparing \acro{} with {\sc MixGreedy}}

Table~\ref{tab:timings} shows the execution times~(columns 2--5), memory usages~(columns 6--8), and influence scores~(columns 9--11) of the baseline algorithm and \acro{} variants. {\sc MixGreedy} runs with a single thread and finishes only in three graphs
{\tt Amazon}, {\tt NetHEP}, and {\tt NetPhy} in $141.3$, $259.1$ and $1725.2$ seconds, respectively. In fact, with a 302,400 seconds~(3.5 days) timeout, these are the only three (out of 12) real-life graphs (with 1.2M, 58.9K, and 231.5K edges) that can be processed by {\sc MixGreedy}. For the others, the original algorithm cannot find a seed set of $K = 50$ vertices within the time limit. However, \acro{} with $\tau = 16$ threads completes all the 12 graphs around 1200 seconds in total, where the maximum runtime is 654.5 seconds for the {\tt Orkut} network having 3.1M vertices and 117.2M edges. The shortest execution time of \acro{} on a graph that cannot be completed by {\sc MixGreedy} is 1.5 seconds. Hence, \acro{} with $\tau = 16$ threads is up to 200,000$\times$ faster than the baseline. 
Only by looking at the sequential execution times of {\sc FusedSampling} on three graphs, we can conclude that $3\times$--$21\times$ of this speedup comes from fusing.  

The fifth column of Table~\ref{tab:timings} presents the execution times of \acro{} to find the first seed vertex which is simply Algorithm~\ref{algo:infusermg} where the {\bf while} loop is executed only once, which is equivalent to the setting with $K = 1$. Comparing these values with the ones in the previous column, we can argue that the benefits of the memoization are more for large $K$ values such as $500$ or $1000$, since most of the time is spent on the {\sc NewGreedyStep-Vec}. For instance, for large graphs, adding the next $49$ seeds only takes $10\%$--$20\%$ of the overall execution time. The actual value depends on the number of the CELF stage; for {\tt Amazon}, to add the remaining seed vertices, \acro{} needs only 79 vertex visits. This is why the cost of the CELF stage is negligible. 

Although it is extremely useful, memoization is also the reason of high memory usage. The values for {\tt NetHEP} and {\tt NetPhy} seem to be relatively lower compared to the baseline. However, these two graphs have only 15K and 37K vertices, much lower than the other graphs. In fact, {\sc FusedSampling} can be a more efficient implementation of {\sc MixGreedy} memory-wise. Comparing the memory use of {\sc FusedSampling} with that of \acro{} reveals the overhead of memoization more clearly. However, even with this overhead, the proposed approach stays practical and extremely efficient on a single server. 

Overall, \acro{} is a practical algorithm, and unlike {\sc MixGreedy}, it can be used on {\em undirected} graphs that have been considered  {\em too large} in the literature. On the  comparable instances, it runs in $2.1$, $0.1$, $0.4$ seconds where {\sc MixGreedy} takes $141.3$, $259.1$, and $1725.2$ seconds, respectively. Furthermore, as the last three columns of Table~\ref{tab:timings} show, the influence scores of the proposed approach are comparable with those of {\sc MixGreedy}.  

\begin{table*} 
\setlength{\tabcolsep}{3pt}
\caption{Memory use~(in GBs) of the algorithms on the networks with $K = 50$ seeds in different simulation settings.}
\label{tab:memoryuse}
\centering
\scalebox{0.90}{
\begin{tabular}{l||r|r|r||r|r|r||r|r|r||r|r|r}
  & \multicolumn{3}{c||}{$p = 0.01$} & \multicolumn{3}{c||}{$p = 0.1$} & \multicolumn{3}{c||}{$p \in N(0.05,0.025)$} & \multicolumn{3}{c}{$p \in [0,0.01]$}\\
  & \multicolumn{1}{c|}{{\sc Imm}} & \multicolumn{1}{c|}{{\sc Imm}}  & \multicolumn{1}{c||}{{\sc InFuseR}} & \multicolumn{1}{c|}{{\sc Imm}} & \multicolumn{1}{c|}{{\sc Imm}}  & \multicolumn{1}{c||}{{\sc InFuseR}} & \multicolumn{1}{c|}{{\sc Imm}} & \multicolumn{1}{c|}{{\sc Imm}}  & \multicolumn{1}{c||}{{\sc InFuseR}} & \multicolumn{1}{c|}{{\sc Imm}} & \multicolumn{1}{c|}{{\sc Imm}}  & \multicolumn{1}{c}{{\sc InFuseR}} \\
  Dataset & \multicolumn{1}{c|}{($\epsilon = 0.13$)} & \multicolumn{1}{c|}{($\epsilon = 0.5$)} & \multicolumn{1}{c||}{{\sc MG}}  & \multicolumn{1}{c|}{($\epsilon = 0.13$)} & \multicolumn{1}{c|}{($\epsilon = 0.5$)} & \multicolumn{1}{c||}{{\sc MG}}  & \multicolumn{1}{c|}{($\epsilon = 0.13$)} & \multicolumn{1}{c|}{($\epsilon = 0.5$)} & \multicolumn{1}{c||}{{\sc MG}} & \multicolumn{1}{c|}{($\epsilon = 0.13$)} & \multicolumn{1}{c|}{($\epsilon = 0.5$)} & \multicolumn{1}{c}{{\sc MG}} \\
    \hline
{\tt Amazon} & 5.46 & 0.55 & 4.05 & 1.76 & 0.24 & 4.05 & 0.82 & 0.16 & 4.06 & 0.82 & 0.16 & 4.06 \\ 
{\tt DBLP} & 5.12 & 0.51 & 6.56 & 10.34 & 1.04 & 6.56 & 2.14 & 0.28 & 6.57 & 2.32 & 0.28 & 6.57 \\ 
{\tt Epinions} & 0.78 & 0.10 & 1.18 & 3.88 & 0.39 & 1.18 & 2.53 & 0.26 & 1.19 & 2.52 & 0.27 & 1.19 \\ 
{\tt LiveJournal} & 71.14 & 9.27 & 75.38 & - & 67.97 & 75.38 & - & 47.35 & 75.38 & - & 47.22 & 75.38 \\ 
{\tt NetHEP} & 0.26 & 0.03 & 0.24 & 0.36 & 0.04 & 0.24 & 0.16 & 0.02 & 0.24 & 0.15 & 0.02 & 0.24 \\ 
{\tt NetPhy} & 0.30 & 0.05 & 0.58 & 1.18 & 0.13 & 0.58 & 0.61 & 0.07 & 0.58 & 0.61 & 0.07 & 0.58 \\ 
{\tt Slashdot0811} & 1.17 & 0.15 & 1.21 & 6.32 & 0.65 & 1.21 & 4.30 & 0.45 & 1.22 & 4.28 & 0.45 & 1.22 \\ 
{\tt Slashdot0902} & 1.22 & 0.16 & 1.29 & 6.67 & 0.69 & 1.29 & 4.53 & 0.47 & 1.30 & 4.50 & 0.47 & 1.30 \\ 
{\tt Orkut} & 172.53 & 20.11 & 50.45 & - & 71.97 & 50.45 & - & 62.93 & 50.45 & - & 62.34 & 50.45 \\ 
{\tt Pokec} & 26.61 & 3.55 & 26.02 & - & 27.13 & 26.02 & 185.55 & 21.27 & 26.22 & 185.54 & 21.02 & 26.22 \\ 
{\tt Twitter} & 2.43 & 0.31 & 1.32 & 10.66 & 1.11 & 1.32 & 8.20 & 0.85 & 1.34 & 8.18 & 0.85 & 1.34 \\ 
{\tt Youtube} & 2.68 & 0.48 & 17.83 & 41.29 & 4.17 & 17.83 & 21.07 & 2.24 & 17.85 & 20.88 & 2.24 & 17.85
\end{tabular}
}
\end{table*}

 \begin{table*} 
\setlength{\tabcolsep}{2.3pt}
\caption{Influence scores of the algorithms on the networks with $K = 50$ seeds in different simulation settings.}
\label{tab:scores}
\centering
\scalebox{0.90}{
\begin{tabular}{l||r|r|r||r|r|r||r|r|r||r|r|r}
  & \multicolumn{3}{c||}{$p = 0.01$} & \multicolumn{3}{c||}{$p = 0.1$} & \multicolumn{3}{c||}{$p \in N(0.05,0.025)$} & \multicolumn{3}{c}{$p \in [0,0.01]$}\\
  & \multicolumn{1}{c|}{{\sc Imm}} & \multicolumn{1}{c|}{{\sc Imm}}  & \multicolumn{1}{c||}{{\sc InFuseR}} & \multicolumn{1}{c|}{{\sc Imm}} & \multicolumn{1}{c|}{{\sc Imm}}  & \multicolumn{1}{c||}{{\sc InFuseR}} & \multicolumn{1}{c|}{{\sc Imm}} & \multicolumn{1}{c|}{{\sc Imm}}  & \multicolumn{1}{c||}{{\sc InFuseR}} & \multicolumn{1}{c|}{{\sc Imm}} & \multicolumn{1}{c|}{{\sc Imm}}  & \multicolumn{1}{c}{{\sc InFuseR}} \\
  Dataset & \multicolumn{1}{c|}{($\epsilon = 0.13$)} & \multicolumn{1}{c|}{($\epsilon = 0.5$)} & \multicolumn{1}{c||}{{\sc MG}}  & \multicolumn{1}{c|}{($\epsilon = 0.13$)} & \multicolumn{1}{c|}{($\epsilon = 0.5$)} & \multicolumn{1}{c||}{{\sc MG}}  & \multicolumn{1}{c|}{($\epsilon = 0.13$)} & \multicolumn{1}{c|}{($\epsilon = 0.5$)} & \multicolumn{1}{c||}{{\sc MG}} & \multicolumn{1}{c|}{($\epsilon = 0.13$)} & \multicolumn{1}{c|}{($\epsilon = 0.5$)} & \multicolumn{1}{c}{{\sc MG}} \\
 \hline
{\tt Amazon}	&	158.5	&	155.5	&	158.6	&	11872.7	&	11743.3	&	12079.8	&	1145.5	&	1129.0	&	1165.5	&	1171.1	&	1174.1	&	1187.9	\\
{\tt DBLP}	&	243.6	&	238.5	&	245.4	&	48550.1	&	48291.5	&	48713.8	&	9967.4	&	9819.3	&	10084.8	&	9774.2	&	9600.3	&	9881.8	\\
{\tt Epinions}	&	3036.7	&	2995.3	&	3051.4	&	20307.7	&	20271.6	&	20362.1	&	12823.8	&	12809.6	&	12871.7	&	12751.5	&	12732.2	&	12793.6	\\
{\tt LiveJournal}	&	260970.1	&	259453.5	&	260364.6	&	-	&	2519467.0	&	2520277.0	&	-	&	1728642.1	&	1729750.3	&	-	&	1723535.8	&	1724181.1	\\
{\tt NetHEP}	&	134.7	&	129.0	&	136.5	&	2462.1	&	2428.0	&	2485.9	&	1118.0	&	1089.2	&	1147.7	&	1116.6	&	1086.8	&	1139.7	\\
{\tt NetPhy}	&	321.5	&	310.3	&	332.5	&	8376.9	&	8271.8	&	8440.0	&	4482.7	&	4436.3	&	4544.8	&	4496.3	&	4417.9	&	4548.2	\\
{\tt Slashdot0811}	&	5166.9	&	5143.4	&	5197.9	&	33446.5	&	33436.3	&	33503.8	&	22428.2	&	22421.6	&	22467.9	&	22361.3	&	22352.7	&	22399.9	\\
{\tt Slashdot090}	&	5399.5	&	5371.0	&	5432.1	&	35127.9	&	35122.0	&	35189.3	&	23466.4	&	23463.7	&	23509.5	&	23406.4	&	23403.7	&	23449.3	\\
{\tt Orkut}	&	650131.4	&	650099.3	&	650237.1	&	-	&	2692357.3	&	2692424.5	&	-	&	2323051.3	&	2323134.3	&	-	&	2320292.3	&	2320340.0	\\
{\tt Pokec}	&	103976.0	&	103906.7	&	104196.3	&	-	&	1096498.0	&	1096625.9	&	835141.2	&	835136.1	&	835258.8	&	833538.6	&	833520.5	&	833643.3	\\
{\tt Twitter}	&	12377.0	&	12294.0	&	12441.6	&	56996.0	&	56926.5	&	57073.1	&	43766.7	&	43675.7	&	43866.8	&	43712.4	&	43559.8	&	43803.6	\\
{\tt Youtube}	&	9130.1	&	8989.4	&	9139.0	&	171362.5	&	171241.8	&	171641.9	&	86582.3	&	86416.4	&	86762.3	&	86196.1	&	86010.9	&	86352.8	
\end{tabular}
}
\end{table*}
 
\subsection{Comparing \acro{} with State-of-the-Art}

To better position \acro{} within the literature, we compare the performance, memory usage, and influence score with a fast, state-of-the-art approximation algorithm {\sc Imm}~\cite{minutoli2019fast} which can produce high-quality seed sets that influences a large number of vertices for both directed and undirected graphs. We also run {\sc Imm} by setting the undirected graph parameter.

\begin{figure*}[htbp] 
    \centering
    \includegraphics[width=0.9\linewidth]{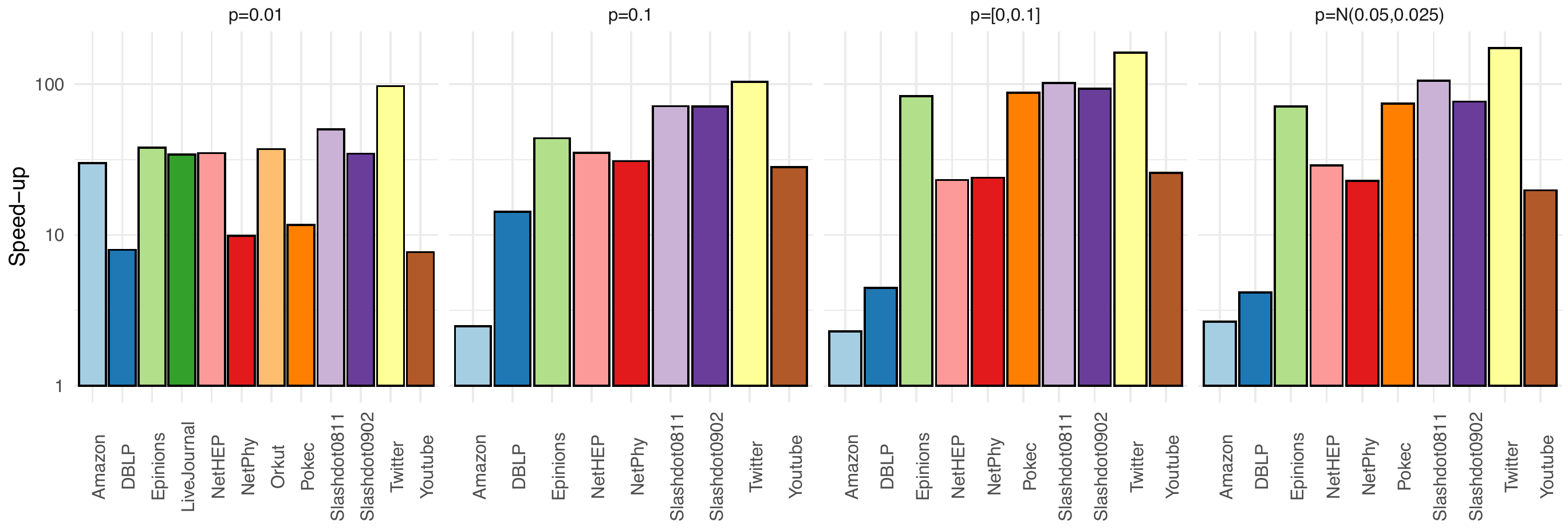}
    \centering\caption{Speedup obtained by \acro{} over {\sc Imm}($\epsilon = 0.13$).
  \label{fig:scaling_vs_imm} }  
\end{figure*}

Tables~\ref{tab:timings-speedup} and~\ref{tab:memoryuse} show the execution times~(in secs.) and memory use~(in GBs), respectively, of \acro{} and two {\sc Imm}  variants for 12 graphs and 4 simulation settings given in Section~\ref{sec:graphs}. The experiments show that \acro{} is $2.3\times$--$173.8\times$ faster than state-of-the-art while always being (marginally) superior in terms of influence scores, and using a comparable amount of memory. As expected, the memory usage of {\sc Imm} is increasing with smaller $\epsilon$ values. In addition, it also increases when the edge weights are larger, i.e., when the samples are denser. For instance, with $p = 0.01$, {\sc Imm}($\epsilon = 0.5$) uses $20$GBs for {\tt Orkut}. However, when $p = 0.1$, the memory usage increases to $72$GBs. Furthermore, {\sc Imm}($\epsilon = 0.13$) cannot run on {\tt LiveJournal}, {\tt Orkut}, and {\tt Pokec} networks due to insufficient memory. On the other hand, \acro{}'s memory usage does not change with different values since it never explicitly creates and stores the samples thanks to fusing. Last, as shown in Table~\ref{tab:scores}, the influence scores of the proposed approach and {\sc Imm}($\epsilon = 0.13$) are comparable. Figure~\ref{fig:scaling_vs_imm} shows the speedup values of \acro with respect to {\sc Imm}($\epsilon = 0.13$).

\subsection{Scalability with multi-threaded parallelism}

\begin{figure*}[!ht] 
    \centering
    \includegraphics[width=0.9\linewidth]{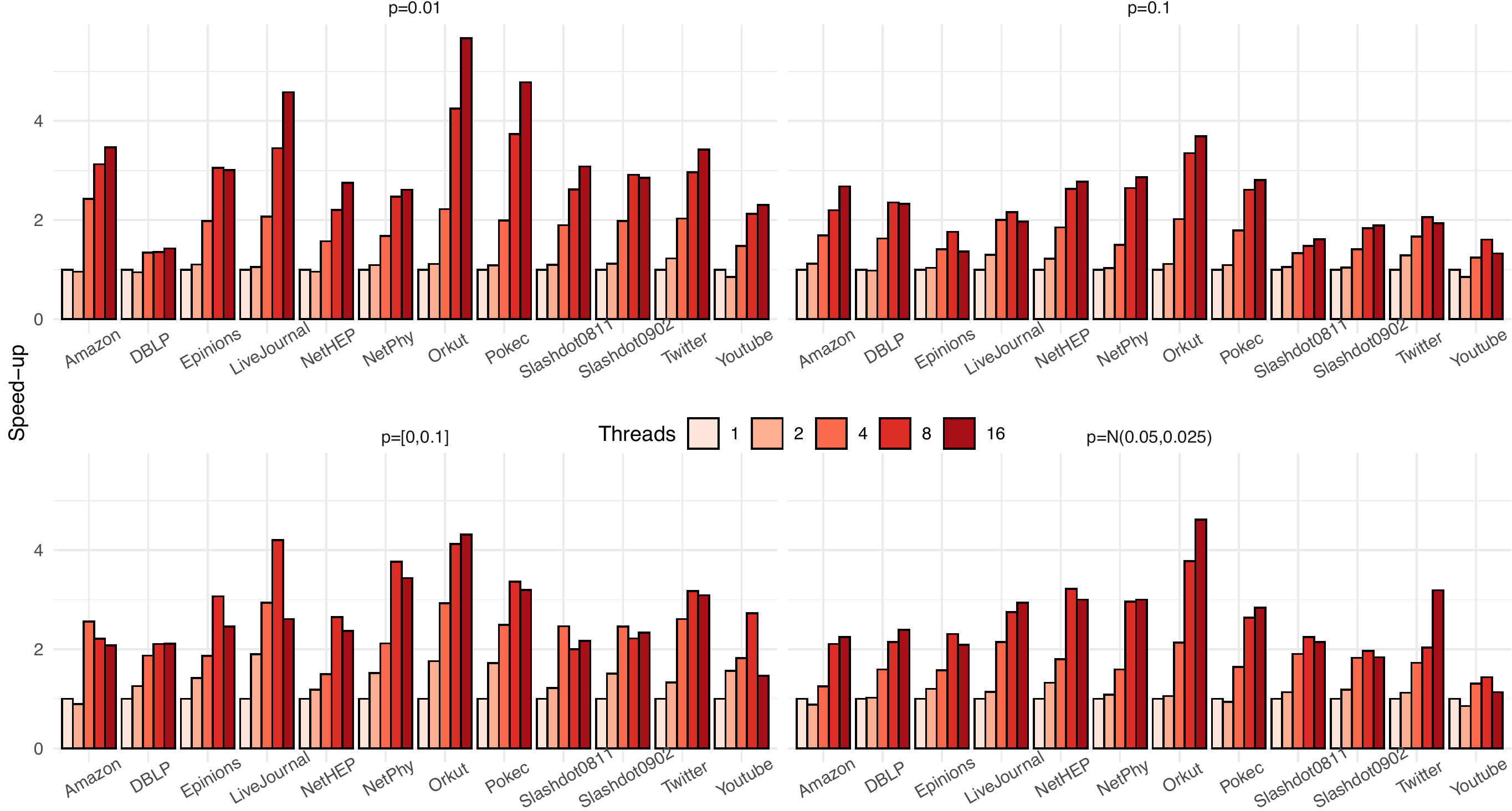}
   \centering \caption{\acro speedup with multiple threads.
    \label{fig:scaling}} 
\end{figure*}

Figure~\ref{fig:scaling} shows the speedup values obtained via {\tt OpenMP} parallelization. 
Since most of time is spent by {\sc NewGreedyStep-Vec}, the parallelization efficiency at line~\ref{ln:for1} of Algorithm~\ref{alg:vecnewgreedy} has a significant impact on the performance. In our implementation,
the parallel processing of live (as source) vertices seems to be necessary to reduce the number of visited edges. However, since a (target) vertex can be a target for multiple sources, the update operation at line~\ref {ln:update} of this {\em push-based} approach is a potential source of race conditions. For denser samples, e.g., for $p = 0.1$, this happens more frequently. Hence, larger influence probabilities may increase (1) the false sharing probability and (2) the number of iterations due to vectorized updates. We argue that these are the reasons for $3\times$--$5\times$ speedup with $\tau = 16$ threads. Still, although the {\em push-based} approach seems necessary, we will investigate {\em pull-based} and {\em hybrid}, i.e., pull/push-based approaches in the future. 

\section{Related Work} \label{sec:rel}

Although they can be inferior in terms of influence score, recent IM algorithms are shown to be quite fast compared to conventional simulation-based approaches such as {\sc MixGreedy}. However, in this work, we show that \acro{}, which is a conventional algorithm, can be orders of magnitude faster than {\sc MixGreedy} while keeping the quality of the seed vertices the same. Techniques such as using GPUs, sketches for finding set intersections, reverse sampling to estimate influence from a small subset of vertices, and estimating the necessary number of simulations/samples required for each step greatly reduces asymptotic boundaries of execution time~\cite{CELF, borgs2014maximizing, minutoli2019fast, SKIM, IPA, IMGPU, curipples}. 

Sketch-based influence maximization improves theoretical efficiency against simulation-based methods. The sketch-based approach pre-computes sketches for evaluating the influence spread instead of running simulations repetitively. One of the interesting methods for sketch-based influence maximization is SKIM~\cite{SKIM} by Cohen~et~al. It constructs bottom-$K^2$ min-hash sketches to estimate the reachability and utilizes multi-core, multi-CPU parallelization. 

Independent Path Algorithm~(IPA)~\cite{IPA} by Kim~et~al runs a proxy model and prunes paths with probabilities less than a given threshold. IPA uses {\tt OpenMP} to work on independent paths in parallel. The approach only keeps a dense but small part of the network and scalable on only sparse networks. Liu~et~al. proposed IMGPU~\cite{IMGPU}, an IM  estimation method by utilizing a bottom-up traversal algorithm. It performs a single Monte-Carlo simulation on many GPU threads to find the reachability of the seed set. It is $5.1\times$ faster than {\sc MixGreedy} on a CPU. The GPU implementation is up to $60\times$ faster with an average speedup of  $24.8\times$.

Borgs~et~al.~\cite{borgs2014maximizing} proposed Reverse Influence Sampling~(RIS) which samples a fraction of all random reverse reachable sets. Then it computes a set of $K$ seeds that covers the maximum number of those. The number of samples is calculated with respect to the number of visited vertices. The algorithm has an approximation guarantee of $(1-1/e-\epsilon)$. Minutoli et al.~improved RIS and proposed {\sc IMM} that works on multi-threaded and distributed architectures~\cite{minutoli2019fast}. 
Recently, the authors extended the algorithm to work on GPUs~\cite{curipples}.

\section{Conclusion and Future Work} \label{sec:conc}
In this work, we proposed fusing and vectorization for IM computations. Better utilization of the CPU cores is achieved by running concurrent simulations at the same time. A comparison with a conventional MC-based algorithm {\sc MixGreedy} and a high-quality, state-of-the-art IM algorithm is presented on real-world datasets and simulation settings. With the proposed techniques, \acro{} can be up to $200000\times$ faster than {\sc MixGreedy} and  $2.3\times$-–$173.8\times$ faster than state-of-the-art on undirected graphs. 

A natural extension of this work is adapting \acro{} to directed graphs. Due to the parallel nature of the simulations, \acro{} can benefit from GPUs if the device memory can be used effectively and efficiently. Also as {\sc Imm}~\cite{minutoli2019fast}, the proposed algorithm can work on larger, massive-scale networks on distributed architectures. In the future, we are planning to pursue these research avenues.

\section*{Acknowledgment}
We would like to thank Dr. Wei Chen for providing the source code and supplementary material of {\sc MixGreedy}. 

\ifCLASSOPTIONcaptionsoff
  \newpage
\fi




\bibliographystyle{IEEEtran}
\bibliography{refs}
%

%

\begin{IEEEbiography}[{\includegraphics[width=1in,height=1.25in,clip,keepaspectratio]{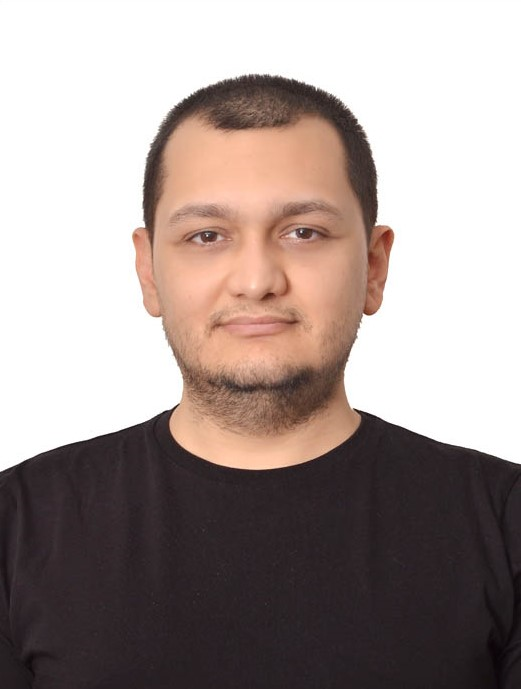}}]{G\"{o}khan~G\"{o}kt\"{u}rk} is a PhD candidate at the Faculty of Engineering and Natural Sciences in Sabancı University. He has received his BS and MS degrees from Sabancı University as well. He is interested in High Performance Computing, Parallel Programming, and Graph Processing.
\end{IEEEbiography}
\begin{IEEEbiography}[{\includegraphics[width=1in,height=1.25in,clip,keepaspectratio]{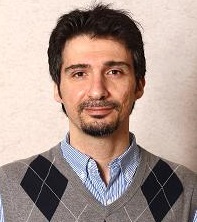}}]{Kamer Kaya} is an Assistant Professor at the Faculty of Engineering and Natural Sciences at Sabancı University. He got his PhD from Dept. Computer Science and Engineering from Bilkent University. He worked at CERFACS, France, as a post-graduate researcher in the Parallel Algorithms Project. He then joined the Ohio State University in September 2011 as a postdoctoral researcher, and in December 2013, he became a Research Assistant Professor in the Dept. of Biomedical Informatics.
His current research interests include Parallel Programming, High Performance Computing, and Cryptography. 
    \end{IEEEbiography}




\end{document}